# The surface degradation and its impact on the magnetic properties of bulk VI$_3$

M. Kratochvílová[1], K. Uhlířová[1], M. Míšek[2], V. Holý[1], J. Zázvorka[1], M. Veis[1], J. Pospíšil[1], S. Son[3,4], J-G. Park[3,4], and V. Sechovský[1]

**AFFILIATIONS**

[1] Charles University, Faculty of Mathematics and Physics, Department of Condensed Matter Physics, Ke Karlovu 5, 121 16 Prague 2, Czech Republic
[2] Institute of Physics, Czech Academy of Sciences, Na Slovance 2, 182 21 Prague 8, Czech Republic
[3] Center for Quantum Materials, Seoul National University, Seoul 08826, Republic of Korea
[4] Department of Physics and Astronomy, Seoul National University, Seoul 08826, Korea

---

**ABSTRACT**

Despite belonging to a well-studied family of transition metal trihalides, VI$_3$ has received significant attention only recently. As a hard ferromagnetic van der Waals compound with a large coercivity, it attracted much attention because of its potential use in atomically thin spintronic and optoelectronic devices. However, practical exploration of VI$_3$ is challenging due to its instability under ambient conditions. Here, we present a comprehensive set of optical, x-ray diffraction, magnetization, and ellipsometric measurements of VI$_3$ and demonstrate that, similarly to the related van der Waals ferromagnet CrI$_3$, the degradation process is accelerated by the presence of moisture. The VI$_3$ surface was covered by selected media commonly used in physical measurements to test its stability and lower the degradation rate three times or higher, providing practical information for experimentalists interested. The decomposition study at ambient conditions shows that the VI$_3$ single crystal can be used for most of the bulk, magnetization, and optical measurements without any noticeable change of physical properties, as the significant degradation appears first after ~ 2 hours of exposition as illustrated, e.g., by the evolution of the ferromagnetic $T_1$ and $T_2$ transitions. The ellipsometric measurement demonstrates that even the surface remains optically stable for at least 5 minutes.

---

## 1. INTRODUCTION

The discovery of a stable magnetic ground state in the monolayer FePS$_3$ in 2016 [1] was an important moment in the recent history of material science, with huge potentials clearly outlined in a separate paper [2]. It was soon followed by a similar discovery in ferromagnetic Cr$_2$Ge$_2$Te$_6$ and CrI$_3$ [3,4]. The family of transition metal trihalides $MX_3$ ($X$ = Cl, B, or I), known for decades, has made a subsequent comeback among the scientific community following these exciting development in recent years. Yet, they are far from a complete understanding, as illustrated by numerous recent experimental and theoretical studies (see, e.g., Soriano [5] and references therein). These materials are interesting concerning their importance in fundamental research and potential applications as spintronic devices based on the existence of two-dimensional (2D) ferromagnetism [5,6,7]. Special attention is given to the new vdW heterostructures that became possible with the appearance of 2D magnets. Nevertheless, the $MX_3$ materials are exciting both in the form of thin-film and bulk [8], as demonstrated in the case of VI$_3$.

VI$_3$ is one of the relatively less studied materials among trihalides [9-14]. Detailed crystal structure study by Doležal et al. [10] has confirmed the *R-3* crystal structure at room temperature; upon cooling, a structural transition into a monoclinic phase is observed at $T_s = 78$ K. There is a consensus that there is bulk ferromagnetism in VI$_3$ at temperatures below $T_C \approx 50$ K [9-14] and the additional magnetic phase transition exists between two ferromagnetic phases at temperatures below 36 K [10,11,15]. Detailed studies of VI$_3$ single crystal revealed two different magnetic phase transitions at $T_1 = 54.5$ K, $T_2 = 53$ K, respectively, observed in ac-susceptibility and low-field magnetization data [11,12,15]; however, no signs were detected in bulk measurements such as specific heat. Based on the analysis of exchange interactions and density-functional-theory (DFT) calculations, these transitions have been attributed to the onset of ferromagnetism in specific crystal-surface layers, which are suffering from iodine deficiency [16] or some lattice defects mimicking intralayer tensile strain. Therefore, the intrinsic bulk ferromagnetism in VI$_3$ exists only at temperatures below the Curie temperature $T_C$. Upon further cooling, another transition at $T_{FM}$ between two different ferromagnetic phases emerges, accompanied by a structure transition from a monoclinic to a triclinic symmetry.

From the practical viewpoint of experimentalists, the nature of the VI$_3$ single crystals and some related di- and trihalides (CrI$_3$ [17,18], Ti- and Zr$X_2$ [19]) reveals a substantial disadvantage – they are metastable at ambient conditions, complicating or even making an impossible number of measurements. The instability of $MX_2$ ($M$ = Cu, Ag) dihalides was studied using the ionic-interaction approach to crystal formation [20], showing that a high ionization potential of the divalent transition metal $M^{2+}$ in combination with large polarizability $\alpha$ of the anion halide (increasing from $\alpha_F \sim 1$ Å$^3$ over $\alpha_{Cl}$ and $\alpha_{Br}$ to $\alpha_I \sim 7.5$ Å$^3$; see [20] and references therein) can explain the non-existence or metastability of some $MX_2$. Analogically, this approach hint at what we can expect in the case of $MX_3$. Indeed, in literature, we can find many studies of trihalides where $X$ = F and Cl. Simultaneously, systems with iodide are less investigated (for review and references, see the work of McGuire [19]). This is nicely illustrated on CrCl$_3$, where the degradation occurs on a time scale at least four orders of magnitude longer than is observed for CrI$_3$ [21]. Also, transition metals with high ionization potential, such as cobalt do not form stable Co$X_3$ to the best of our knowledge [22].

It has therefore become crucial for future experiments to know enough of the actual sensitivity of the materials to various environments and their abilities to protect the sample from degradation and the time scales, in which the single crystal substantially degrades. Also, understanding the degradation process and its outcomes is necessary for the interpretation of the measurement results. As was already mentioned, the VI$_3$ surface degradation can cause the formation of surface iodine-poor layers, which may become ferromagnetic at temperatures higher than $T_C$ of the bulk. In the case of VI$_3$ and related materials unstable at ambient conditions, it might be very useful to thoroughly characterize their "instability" upon actual experimental conditions that occur in the standard physical measurements, such as x-ray diffraction or hydrostatic-pressure experiments. Therefore, we have investigated the time degradation of the VI$_3$ single crystal surface by taking optical images of comparable samples surrounded by the Daphne Oil 7373 [23] utilized routinely in high-pressure measurements, a perfluoropolyether lubricant (PFPE) used in single-crystal x-ray diffraction for comfortable separation and selection of single crystals without showing any diffraction peaks at low temperatures, and the vacuum oil (Edwards ULTRAGRADE™ Performance 19 Oil) used as an example of a medium for possibly stable storage of the $MX_3$ single crystals. It should be noted that the pressure medium freezes above 2.2 GPa at room temperature [24], which effectively changes the environment of the crystal and therefore affects its stability. Except for environments with a generally protective effect, we have also studied an environment that accelerates the sample degradation - demineralized water - to illustrate the corroding effect of air moisture. As was pointed out in the study of few-layer CrI$_3$ sheets [18], another crucial factor affecting the samples' stability is the light exposure. Therefore, we have performed a comparable set of experiments in the dark to check the effect of light exposure on surface degradation. We have measured the temperature dependence of magnetization after defining exposition to the air to characterize the degradation products quantitatively. We have analyzed the surface degradation effects on the $T_1$ and $T_2$ transitions being ascribed to the surface quality.

## 2. EXPERIMENTAL DETAILS

The $VI_3$ single crystals were grown by chemical vapor transport method according to the procedure described by Son et al. [9]. The samples had a shape typical for vdW materials, thin plates with some hexagonal-like edges, and the *c*-axis perpendicular to the plates at room temperature. Their typical lateral dimensions were up to several millimeters. The samples were stored under an inert Ar (6N purity) atmosphere in a glovebox with long-term stability either in the dark or exposed to light. Except for the defined exposure time, the contact of the samples with air was minimized only at the time of the installation on the sample holders and in the cryostat, which took less than 15 minutes in all cases. A fresh sample was used for every separate measurement unless stated otherwise. We omitted samples with many cracks or deformations (except the diffraction experiment as commented below) as they are generally a source of degradation. Hence, they can affect the degradation rate.

A MIRA3-FEG scanning electron microscope (SEM) equipped with Bruker Flash energy-dispersive X-ray (EDS) analyzer was used to confirm the stoichiometry and phase purity of the samples. The samples were characterized by powder x-ray diffraction (Powder Diffractometer Bruker D8 Advanced equipped with monochromatic CuKα radiation and Rigaku SmartLab). High-resolution XRD using Rigaku SmartLab (CuKα rotation-anode generator operated at 45 kV/200 mA) in a parallel beam mode was performed on an as-grown single crystal oriented in the (001) direction. Magnetization data were obtained utilizing MPMS XL 7T (*Quantum Design Inc.*) using a crystal oriented with the *c\**-axis (≡ the surface normal) parallel to the applied magnetic field. The measurement was performed in the zero-field-cooled mode (ZFC). The optical response measured in the photon energy range from 0.7 to 6.5 eV using a Woollam RC2 ellipsometer equipped with light focusing probes at various incidence angles.

## 3. RESULTS AND DISCUSSION

### 3.1. Stability of $VI_3$ under Various Environments

For the study of the stability of $VI_3$ single crystals with similar habitus were chosen. The single crystals have preferably 1 mm lateral dimension with precise growth facets and without cracks or deformations unless specified otherwise. The following method of sample handling before the experiments was applied: first of all, the single crystal was cleaved using sticky tape within the Ar glovebox. After that, its surface was covered by the selected protective layer. The sample was always placed in the middle of a ring filled by the protective medium, as shown in Fig. 1f, forming a bath to ensure that the sample's surface is covered during the whole measurement. After this procedure, the samples were taken out of the glove box and placed under the optical microscope to monitor the time-dependent degradation. The samples were exposed to light from the optical microscope during the whole measurement except the experiments in the dark, where the samples were illuminated only during photo capture and stored in a dark box otherwise.

The time dependence of sample degradation was captured for several days until the sample was not fully decomposed, i.e., only the droplet of the iodine water solution remained. The highest impact on the decomposition time had the sample environment. However, every single crystal's size and starting condition also played a particular role, which cannot be quantitatively evaluated. Experiments with the same conditions were repeated with several samples with good qualitative reproducibility (the total time of degradation could differ with a factor of 2 depending on sample quality). Nevertheless, a good understanding of the impact of various environments on VI$_3$ can be obtained.

The degradation is induced and propagated based on intrinsic defects, such as grain boundaries and point defects, because of their high chemical reactivity due to lattice breakage. Higher densities of dislocations and point defects lead naturally to higher chemical reactivity and faster decomposition. Notably, the grain boundaries in VI$_3$ exhibit degraded edges with triangular and hexagonal shapes, which are well observable due to the droplet formation on the edges, as illustrated in Fig. 1. The mechanism of iodide droplet formation on the sample's surface seems similar at ambient conditions and under-tested protective layers. First, tiny droplets form at the grain boundaries (see Fig. 1e) and crystal edges, and other defects, and they grow continuously. Later, they merge into microscopic droplets that usually spread from the sample's edges and finally they cover the majority of the crystal's surface. Thus, in line with the previous study of the CrI$_3$ degradation [18], we identify water as the primary reactant with VI$_3$ under ambient conditions. The I$^-$ ligand on the sample's surface is substituted by water leading to a photocatalytic formation of aquavanadium complexes and dissociation of VI$_{3-x}$(H$_2$O)$_x^{x+}$ and I$^-$ ions on the samples' surface. This process escalates the hygroscopy and further dissolution; once these droplets are dissolved in demineralized water, they leave no visible residue.

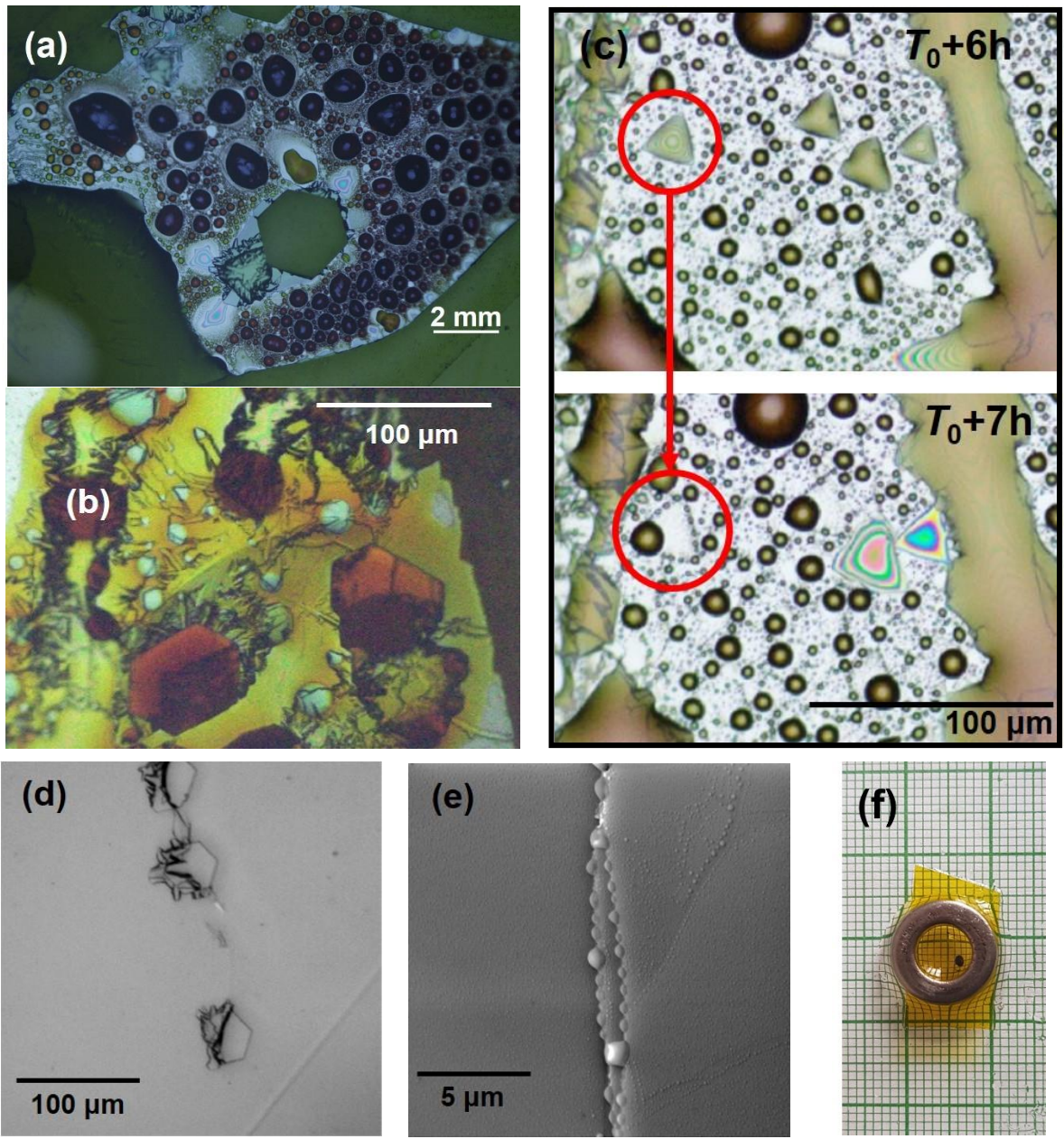

**Fig. 1.** a) a hexagonal-shaped pit formed in the degraded grain in the PFPE Oil (~$T_0$+2 days), b) hexagonal-shaped pits filled with iodine solution in the sample covered by the Daphne 7373 Oil (~$T_0$+10 hours), c) a triangular-shaped grain covered by a droplet stretched over its due to surface tension at ambient conditions at $T_0$+6 hours and below the exact grain after one more hour with tiny droplets sitting on its boundaries and the large droplet formed in one of its corners, d) hexagonal-shaped facets formed by demineralized water corroding the surface (~$T_0$+15 minutes), e) the surface from SEM showing tiny submicron holes remaining after the droplets evaporated (~$T_0$), and f) sample placed in a ring filled by the protective medium forming a bath.

The degradation pathways in different environments are summarized in Fig. 2 and 3, showing the comparison of degraded surfaces under various conditions in different time scales. As could be expected, all selected types of oils slow down the degradation as compared to ambient conditions by protecting the sample from the air moisture (see Fig. 2). The remains of the sample are entirely covered by a droplet of the iodine water solution after one day at ambient conditions. Simultaneously, we can still observe the contours of the crystal and new surface for the other cases. The differences among degraded surfaces in these experiments are not significant except for the vacuum oil. The vacuum oil keeps the majority of the VI$_3$ surface untouched for more than three days. The surface of other samples is completely decomposed after this time, prolonging the protection using the vacuum oil by a factor of 3. The complete surface decomposition of the sample immersed in the vacuum oil was observed first after nine days (not shown). The sample covered by a droplet of demineralized water degrades entirely in two hours, which is more than ten times faster than the degradation of a sample kept at ambient conditions. From the practical point of view, the degradation monitoring under the PFPE oil showed that it could be used safely for most diffraction measurements, which take several hours: even after which the sample remains stable. Also, the Daphne oils used in the pressure measurements protect the sample reasonably well if we consider usual experimental conditions, i.e., low temperatures, which further reduce the kinetics of chemical reactions.

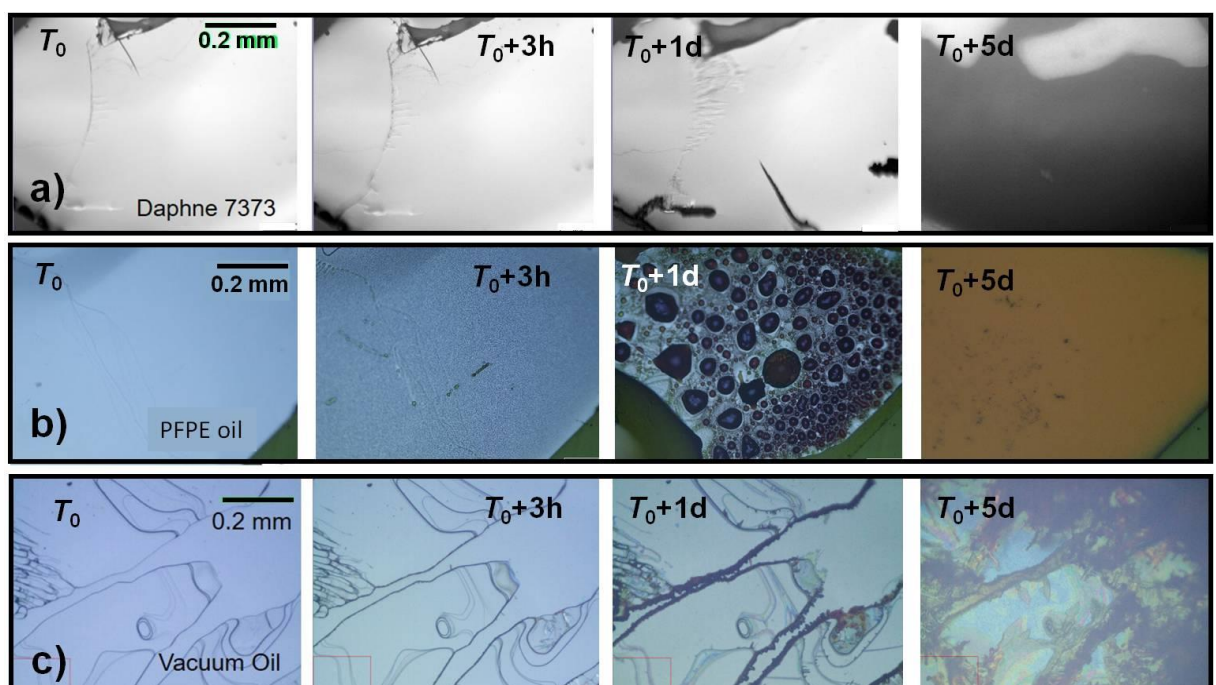

**Fig. 2.** The time evolution of the degradation of the VI$_3$ surface covered by various protective layers: a) Daphne Oil 7373, b) PFPE Oil, and c) Vacuum Oil, observed under optical microscope up to 5 days.

Fig. 3 shows the difference between the degradation processes taking place in the dark and illuminated with light. The VI$_3$ surface is long-term stable in argon independently on the illumination and in the air; the situation does not seem to be dramatically different; here, the sample is completely dissolved after one day when illuminated and after ~2 days when kept in the dark. Also, the degradation rate of VI$_3$ in demineralized water (see Fig. 3c and d) is somewhat comparable for both light conditions, reflecting the robustness of the reactivity with water, which cannot be slowed down by the dark noticeably. In general, the differences between the degradation processes in the air are not in orders of magnitude as in CrI$_3$ [18], as two days of the stability difference correspond roughly to the experimental error. Thus, we cannot say that the illumination of the VI$_3$ surface accelerates the degradation similarly to CrI$_3$, but the reason might be that we study bulk samples and not few-layer flakes, magnifying the different conditions of the samples' environment.

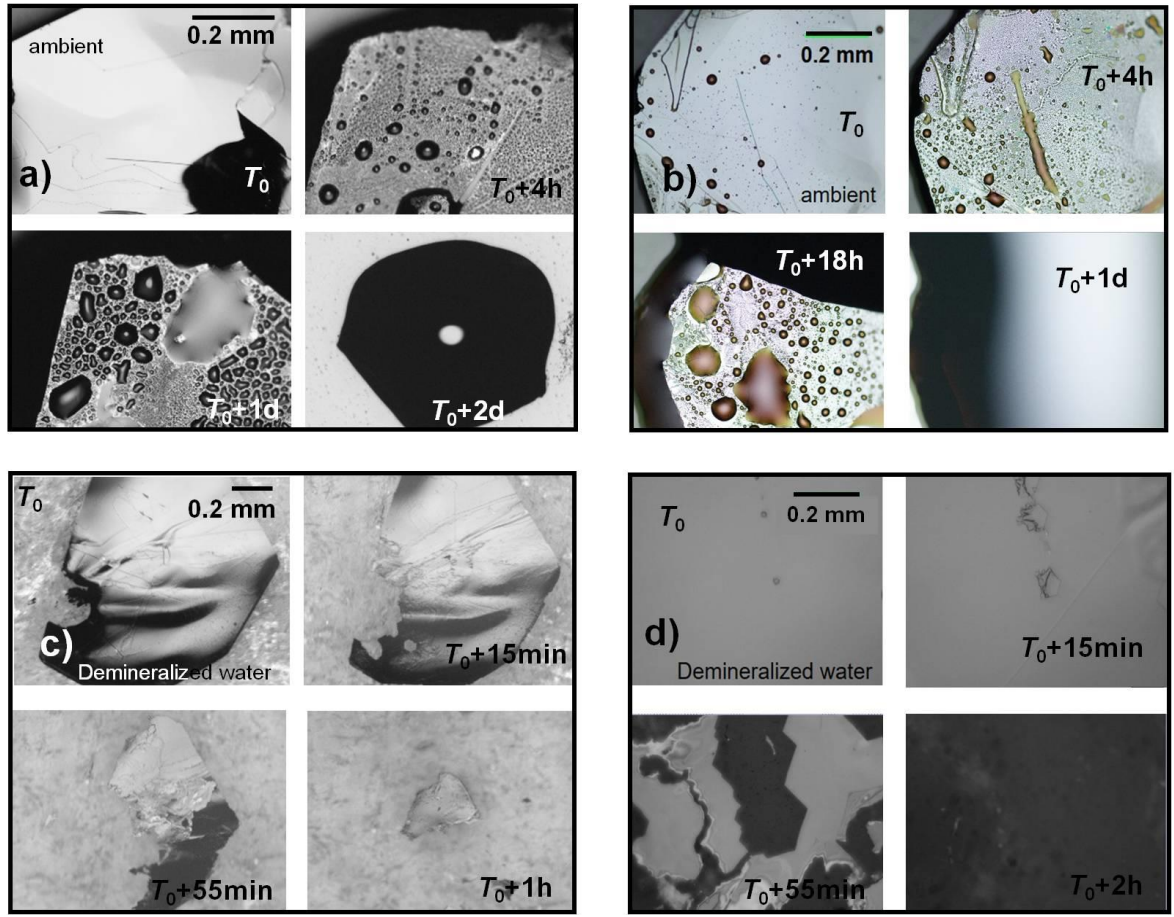

**Fig. 3.** The time evolution of the degradation of the VI$_3$ surface a) in the air in the dark, b) in the air illuminated with light, c) immersed in demineralized water in the dark, and d) ) immersed in demineralized water illuminated with light.

*3.2. XRD measurement*

We have investigated the time evolution of the XRD diffraction pattern *in situ* at ambient conditions and repeated the measurement several times to ensure reproducibility. Two types of experiments were performed: (#1) one single crystal was placed on a non-diffracting silicon plate, and the symmetric $2\theta/\omega$ scans around the (00l) reflection maxima were measured in 20-minute cycles up to ~ 17 hours; (#2) single crystals were cracked into pieces made as small as possible, i.e., ~300 µm dimensions (standard pulverization could not be realized because of the softness and minimal thickness of the layered vdW crystals). After that, they were placed on a non-diffracting silicon plate, and the diffraction patterns were measured in ~40-minute cycles up to 4 days and then again after one and two weeks approximately.

The significant preferential orientation shown in Fig. 4a has an origin in the layered structure of VI$_3$. New peaks in both types of experiments start emerging after ~ 2 h while the peaks belonging to VI$_3$ gradually decay. Integrated intensities of the (00*l*) peaks from measurement #1 show different time evolution (see Fig. 4b), although one might naively expect that they would decay similarly due to the decreasing amount of VI$_3$. However, as the sample dissolves in a droplet of the iodine water solution, it moves gradually from its original position/orientation, and sooner or later, it stops fulfilling the diffraction condition; thus, the degradation disables the *in situ* measurement as it affects the intensities' evolution of the VI$_3$ diffraction peaks. On the other hand, crushing the samples before the XRD measurement done in experiment #2 does not lead to significant differences between the (00*l*) intensity-decays as shown in the inset of Fig. 4b.

The resulting product of the decomposition was stable after one and more weeks exposed to air, i.e., the crystals had a similar habitus and composition as was shown via the microprobe (EDS) analysis and SEM (see inset in Fig. 4a). Needle-shaped single crystals of typical length ~ 50 μm were formed. According to the microprobe, their stoichiometry varies from V:I ~1:1 to ~2:1, pointing clearly to the lack of iodine. As a light element, Oxygen could not be determined via this method quantitatively; however, its presence in $VI_3$ is confirmed as it is easily detectable in the EDS spectra of the crystals' surface compared to spectra measured elsewhere. The resulting degradation product on the silicon plate exposed to x-rays contrasts sharply with the one left on a glass plate or ampoule either at ambient conditions or covered by various protective layers. This means that the structure and stoichiometry of these crystals exposed to x-rays do not correspond to the samples subjected to most of the standard measurements. From this point of view, the XRD analysis is not a very useful tool as it directly interacts with the $VI_3$ single crystals, affecting the degradation process.

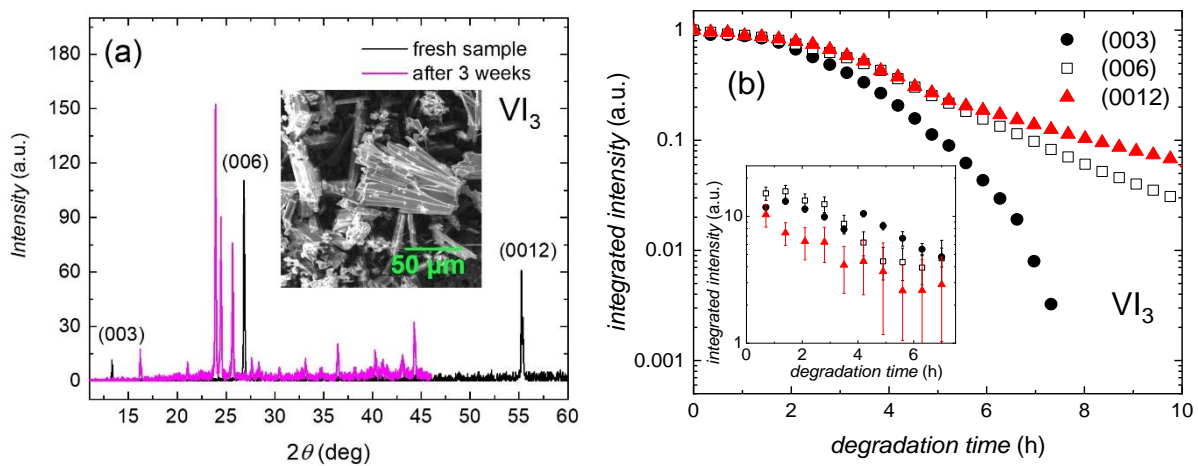

**Fig. 4.** a) The diffraction pattern of a fresh sample of $VI_3$ (black) and after approximately three weeks at ambient conditions (magenta). The fresh sample pattern shows the (00*l*) peaks only due to a strong preferential orientation of the vdW layers. The inset shows an SEM picture of a degradation product of the pulverized $VI_3$ single crystals placed on a silicon diffraction plate after exposure to ambient conditions for more than one week. The image was obtained using the secondary electron detector. b) The time evolution of the (00*l*) peaks belonging to the $VI_3$ phase from experiment #1. The inset shows analogical data from experiment #2.

*3.3. DC magnetization studies*

To trace the evolution of the non-intrinsic transitions $T_1$ and $T_2$ upon increasing surface degradation induced by exposure to ambient conditions, the temperature-dependent low-field magnetization $M(T)$ was performed. The single crystals were aligned within the *ab* basal plane in the magnetic field of 10 mT, and they were measured as fresh samples, samples after 1 hour, 5 hours, and 27 hours. After 27 hours, the sample was mechanically cleaned from the droplet of the iodine water solution and measured immediately again (marked as "cleaned" in Fig. 5). The solid pink line depicts the measurement of the sample after 27 hours at ambient conditions, while the pink dashed line marks the measurement of the identical, cleaned sample. The absolute values of magnetization below $T_C$ ~ 50 K in Fig. 5a and $M(\mu_0 H=0)$ shown in the inset of Fig. 5b decrease upon increasing exposure time. The decrease reflects the decreasing mass of the non-degraded ferromagnetic VI$_3$ and a simultaneous increase of the degradation product in the paramagnetic liquid form. For simplicity, assuming that the absolute value of remanent magnetization $M(0)$ corresponds directly to the ferromagnetic/paramagnetic material ratio, we realize that after one day of exposure, less than ~ 10% of the original amount of the VI$_3$ sample is left. This bulk information is in line with the "2D" monitoring of the samples' surface from Fig. 3b.

Looking at the magnetization curves measured after the 27 h-exposure with and without the formed droplet, respectively, one can see that the saturation magnetization is approximately ten times lower for the "cleaned" sample, which is caused by the paramagnetic contribution of the droplet. On the other hand, the remnant magnetization is almost untouched by the degradation time comparing the new and "cleaned" sample data, in agreement with previous observations. The coercive field values are strongly influenced by the amount of the paramagnetic contribution from the droplet.

Fig. 5c shows the first derivative - $\partial M/\partial T$ in the narrow temperature region above $T_C$, where the maxima corresponding to the transitions $T_1$ and $T_2$ appear. While the ferromagnetic temperature does not change with increasing degradation noticeably, this is not the case of $T_1$ and $T_2$. Plotting their values versus the exposure time, we can observe a slight increase of both $T_1$ and $T_2$ in Fig. 5d; removing the droplet of the iodine water solution after the 27-hour exposure partially restores the transition temperature values of the fresh sample. According to the scenario where the iodine deficiency in specific-surface layers causes the increase of ferromagnetic temperature, one might suggest that with longer degradation time, the deficit of iodine propagates into the bulk creating more such deficient layers. These layers would manifest themselves with additional bumps on the - $\partial M/\partial T$ curves, i.e., showing other ferromagnetic transitions. This suggestion cannot be confirmed by our data as already the bump corresponding to the $T_1$ transition is difficult to identify in the degraded sample. Instead, more prolonged exposure to ambient conditions leads to a slight increase of $T_1$ and $T_2$ within the two surface layers. Also, the evolution of the peak area probably reflects that a modest exposure increases the transition temperatures within the iodine-deficient layers. However, with advancing degradation, the peak smears out as the material decomposition starts to dominate. It seems that the propagation stops probably after only several iodine-deficient layers are established as further air exposure leads to complete decomposition of VI$_3$ and now to the loss of ferromagnetic properties.

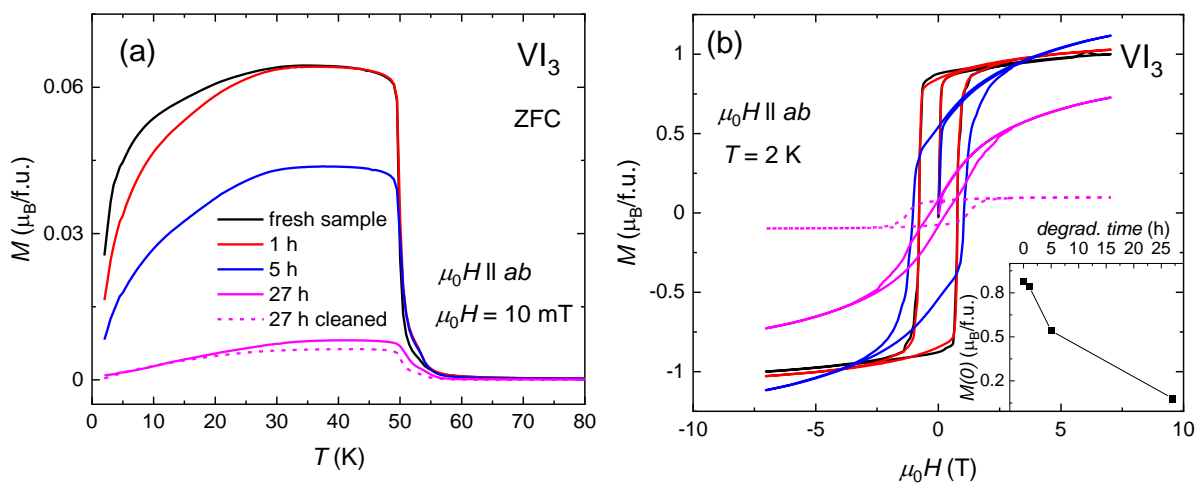

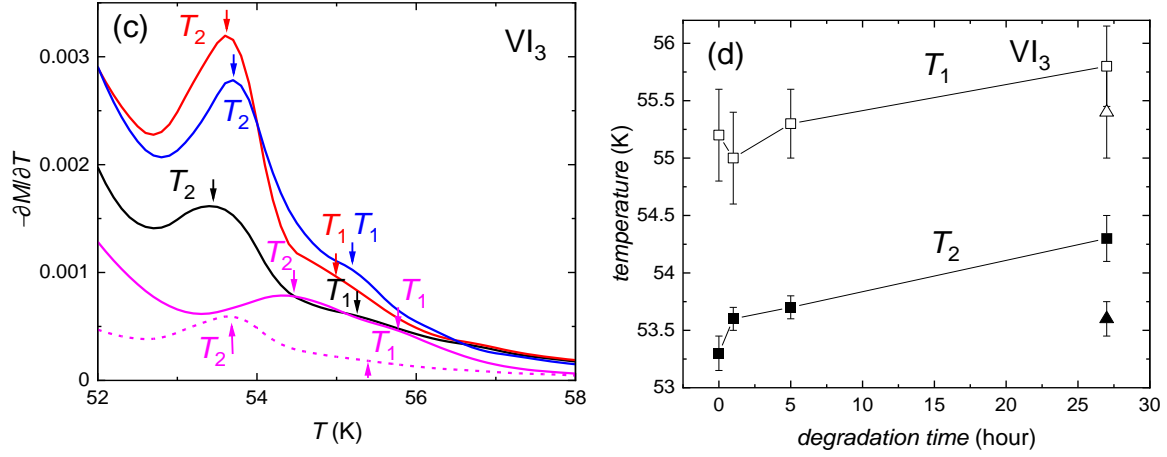

**Fig. 5.** a) The temperature-dependent ZFC magnetization of VI$_3$ for various states of sample degradation measured at 10 mT aligned within the *ab* basal plane. The legend from a) is identical for data shown in b) and c), respectively. b) The respective hysteresis curves measured at 2 K (the inset shows the time evolution of *M*(0)), c) the $-\partial M/\partial T$ vs. *T* plot illustrating the positions of the $T_1$- and $T_2$-anomalies in VI$_3$ being exposed to the ambient environment for various exposure times (the arrows mark the maxima of the peaks, i.e., the positions of the $T_1$- and $T_2$-temperatures), and d) the time-dependent evolution of the transition temperatures $T_1$ and $T_2$. A triangle marks the data for the cleaned sample for both time evolutions. The lines are guides for the eye.

*3.4. Optical studies at ambient atmosphere*

Since the optical measurements are surface sensitive for energies higher than the bandgap of a particular material, we employed spectroscopic ellipsometry to observe optical response changes with time. For this purpose, ellipsometric variables $\Psi$ and $\Delta$ defined by the ratio of amplitude reflection coefficients for *s* and *p* polarized light ($r_p/r_s = \tan\Psi e^{i\Delta}$) were acquired in a broad spectral range. Several mm wide light spot covered the whole surface area of the measured crystal and the spectra of $\Psi$ and $\Delta$ were recorded for one minute at the angle of incidence 60 degrees. The measurement was repeated at selected time intervals from the exfoliation of the sample surface. The resulting spectra of $\Psi$ are shown in Fig. 6a. For the first 95 minutes, the spectral behavior remains almost constant, and the spectra are only shifted with respect to each other; a notable change occurred 185 minutes after the exfoliation. The significant shift towards lower values and an appearance of additional spectral structure around 5.3 eV are visible.

Moreover, a several hundred nanometers thick surface layer of different material, which is transparent up to 3.5 eV, revealed itself as a substantial interference at the lower energy side in the spectrum. The spectra were taken at 240 and 340 minutes after the exfoliation demonstrate complete suppression of VI$_3$ response on the surface and an onset of a material with a completely different optical response with prominent spectral structures in the UV region. This material seems to be relatively stable from the optical point of view for the last 100 minutes of the measurement. Inset of Fig. 6a demonstrates the change of $\Psi$ at the energy of 3.1 eV within the whole time window. It is visible that after a relatively stabilized phase, the main change appears between 2 and 4 hours after the exfoliation, followed by another stabilized phase. The results are consistent with the XRD results shown in Fig. 4b and with the ZFC magnetization in Fig. 5a.

Since the light irradiation by 3 mm wide macroscopic spot covered the whole surface of the investigated sample containing many degradation centers (i.e., grain boundaries, cracks, and other defects), the second set of measurements was done employing the same ellipsometer, but now with focusing probes. The spot of about 200 micrometers was focused at a 65-degree incident angle to a large plane facet of the sample's surface with minimal defects using a microscope camera. The measurements were taken mainly within the first 20 minutes in a broader spectral range to observe subtle changes in optical response after the exfoliation. The resulting spectra of $\Psi$ are shown in Fig. 6b. A rapid change of spectral behavior was observed for all measured times at around 0.81 eV. This indicates bandgap energy where the sample becomes transparent for lower energies. The estimated value reasonably agrees with values obtained by ARPES (approx. 1 eV [25], DFT calculation (0.98 eV [26]) and reflectance measurements (0.6 eV [13]). One has to note that our estimation of the energy (marked in Fig. 6b by an arrow and $E_g$) is very roughly based on the spectral change and not deduced by numerical analysis of the data since this is not the purpose of this paper. The time evolution of the spectra in Fig. 6b demonstrates that during the first 6 minutes after the exfoliation one can consider the sample optically stable within the measurement error and very subtle changes. A nanometer-thick surface layer is probably evolving during this period but not noticeably affecting the overall optical response. After this first evolution stage, a notable change in the spectral behavior is obtained, combining a gradual shift and spectral modulation. This suggests a further development in the thickness and the composition of the surface layer.

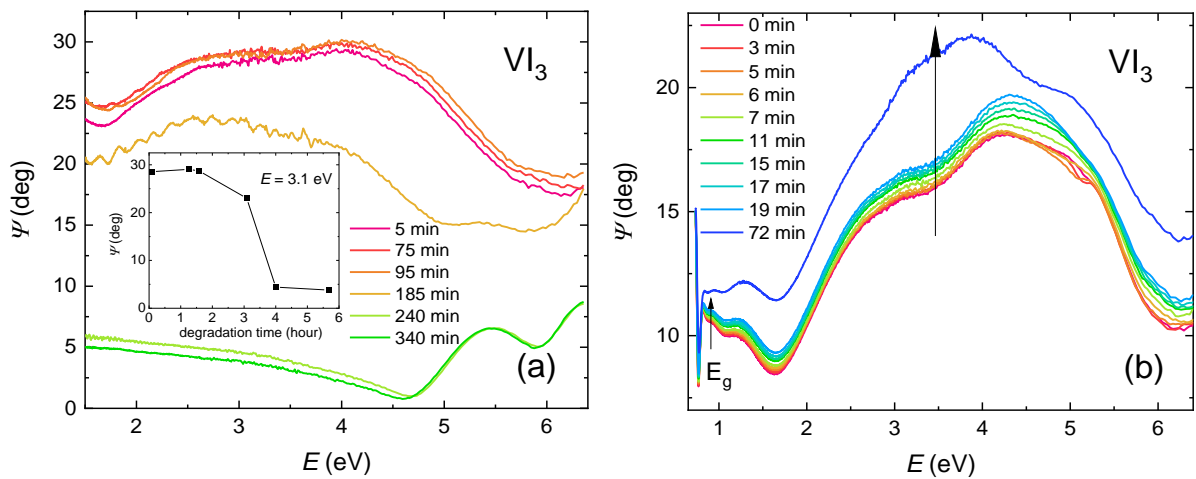

**Fig. 6.** a) Time evolution of $\Psi$ spectra with respect to the exfoliation. The inset shows $\Psi$ vs. time for the energy of incident light $E = 3.1$ eV. The whole surface of the sample was irradiated. b) Time evolution of $\Psi$ spectra with respect to the exfoliation. An arrow and $E_g$ mark a notable change of the spectral behavior due to the bandgap energy. The large arrow marks the time flow. The sample was irradiated using focusing probes, and the spot size was 200 micrometers.

## 4. SUMMARY AND CONCLUSIONS

In summary, we demonstrate that the oil media such as the Daphne 7373 oil, the PFPE oil, and especially the vacuum oil help maintain the sample's surface clean from moisture, thereby protecting the sample from ambient conditions slowing down the degradation rate by a factor of 3 or higher. We were not able to confirm sensitivity to light exposure due to the bulk nature of our samples compared to the $CrI_3$ flakes [18]. The corrosion of the $VI_3$ single crystals studied by x-ray powder diffraction illustrates the gradual decay of the trigonal layered vdW structure and an emergence of a different surface layer after a ~2-hour exposition; however, the interaction of the $VI_3$ degradation products with the high-energy light source and the silicon plate does not correspond to the conditions and thus, the degradation process, observed in all other types of measurements. We can see some deviation at low temperature between the fresh and 1h-exposed sample in the ZFC temperature dependence of magnetization. The bulk samples' magnetic properties are well preserved up to ~ 5 hours being exposed to ambient conditions.

One-day exposure leads to the slight increase of $T_1$ and $T_2$ transition temperatures within the two surface layers while the ferromagnetic temperature $T_C$ stays the same. After that, in line with the optical monitoring, the sample of $VI_3$ is almost entirely dissolved, leading to the loss of ferromagnetic properties. Although the sample degradation is relatively fast, we can conclude that within the usual time needed to install the sample to the measurement holder, the sample's magnetization does not change significantly. Still, the high paramagnetic signal from the decomposition product should be taken into account when comparing measurements from different samples. The ellipsometric study showed that the surface of $VI_3$ is relatively optically stable for the first 6 minutes of ambient conditions exposure. Moreover, the stability of the whole $VI_3$ single crystal extends to approx. 2 hours before degrading into a different phase, being consistent with the XRD, magnetization, and optical experiments.

## ACKNOWLEDGMENTS

This work is part of the research program GACR 19-16389J which is financed by the Czech Science Foundation. Works at SNU were supported by the Leading Researchers Program of the National Research Foundation of Korea [NRF-2020R1C1C1013642] of the Republic of Korea. This project was supported by OP VVV project MATFUN under Grant No. CZ.02.1.01/0.0/0.0/15_003/0000487. Experiments were performed in MGML (https://mgml.eu/), which is supported within the program of Czech Research Infrastructures (project no. LM2018096).

## DATA AVAILABILITY

The data that support the findings of this study are available within the article.

## REFERENCES

[1] J. Lee, S. Lee, J. Ryoo, S. Kang, T. Kim, P. Kim, Ch. Park, J-G. Park, H. Cheong, Ising-Type Magnetic Ordering in Atomically Thin $FePS_3$, Nano Lett. 16 (2016) 7433.

[2] J-G. Park, Opportunities and challenges of 2D magnetic van der Waals materials: magnetic graphene? J. Condens. Matter Phys. 28 (2016) 301001.

[3] Ch. Gong, L. Li, Z. Li, H. Ji, A. Stern, Y. Xia, T. Cao, W. Bao, Ch. Wang, Y. Wang, Z. Q. Qiu, R. J. Cava, S. G. Louie, J. Xia, X. Zhang, Discovery of intrinsic ferromagnetism in two-dimensional van der Waals crystals, Nat. 546 (2017) 265–269.

[4] B. Huang, G. Clark, E. Navarro-Moratalla, D. R. Klein, R. Cheng, K. L. Seyler, D. Zhong, E. Schmidgall, M. A. McGuire, D. H. Cobden, W. Yao, D. Xiao, P. Jarillo-Herrero, X. Xu, Layer-dependent ferromagnetism in a van der Waals crystal down to the monolayer limit, Nat. 546 (2017) 270–273.

[5] D. Soriano, M. I. Katsnelson, J. Fernández-Rossier, Magnetic Two-Dimensional Chromium Trihalides: A Theoretical Perspective, Nano Lett. 20 (2020) 6225–6234.

[6] M. Gibertini, M. Koperski, A. F. Morpurgo, K. S. Novoselov, Magnetic 2D materials and heterostructures, Nat. Nanotechnol. 14 (2019) 408–419.

[7] Ch. Gong, X. Zhang, Two-dimensional magnetic crystals and emergent heterostructure devices, Science 363 (2019) 6428.

[8] M. Gibertini, Magnetism and stability of all primitive stacking patterns in bilayer chromium trihalides, J. Phys. D: Appl. Phys. 54 (2021) 064002.

[9] S. Son, M. J. Coak, N. Lee, J. Kim, T. Y. Kim, H. Hamidov, H. Cho, C. Liu, D. M. Jarvis, P. A. C. Brown, J. H. Kim, C. H. Park, D. I. Khomskii, S. S. Saxena, J. G. Park, Bulk properties of the van der Waals hard ferromagnet $VI_3$, Phys. Rev. B 99 (2019) 041402(R).

[10] P. Doležal, M. Kratochvílová, V. Holý, P. Čermák, V. Sechovský, M. Dušek, M. Míšek, T. Chakraborty, Y. Noda, S. Son, J. G. Park, Crystal structures and phase transitions of the van der Waals ferromagnet $VI_3$, Phys. Rev. Mater. 3 (2019) 121401(R).

[11] J. Valenta, M. Kratochvílová, M. Míšek, K. Carva, J. Kaštil, P. Doležal, P. Opletal, P. Čermák, P. Proschek, K. Uhlířová, J. Prchal, M. J. Coak, S. Son, J-G. Park, V. Sechovský, Pressure-induced large increase of Curie temperature of the van der Waals ferromagnet $VI_3$, Phys. Rev. B 103 (2021) 054424.

[12] Y. Liu, M. Abeykoon, C. Petrovic, Critical behavior and magnetocaloric effect in $VI_3$, Phys. Rev. Res. 2 (2020) 013013.

[13] T. Kong, K. Stolze, E. I. Timmons, J. Tao, D. R. Ni, S. Guo, Z. Yang, R. Prozorov, R. J. Cava, $VI_3$—a New Layered Ferromagnetic Semiconductor, Adv. Mater. 31 (2019) 1808074.

[14] S. J. Tian, J. F. Zhang, C. H. Li, T. P. Ying, S. Y. Li, X. Zhang, K. Liu, H. C. Lei, Ferromagnetic van der Waals Crystal $VI_3$, J. Am. Chem. Soc. 141 (2019) 5326.

[15] E. Gati, Y. Inagaki, T. Kong, R. J. Cava, Y. Furukawa, P. C. Canfield, S. L. Bud'ko, Multiple ferromagnetic transitions and structural distortion in the van der Waals ferromagnet $VI_3$ at ambient and finite pressures, Phys. Rev. B 100 (2019) 094408.

[16] J. Yang, J. Wang, R. Liu, Q. Xu, Y. Li, M. Xia, Z. Li, F. Gao, Enhancement of ferromagnetism for $VI_3$ monolayer, Appl. Surf. Sci. 524, (2020) 146490.

[17] L. L. Handy, N. W Gregory, A study of the chromous-chromic iodide equilibrium, J. Am. Chem. Soc. 70 (1950) 5049.

[18] D. Shcherbakov, P. Stepanov, D. Weber, Y. Wang, J. Hu, Y. Zhu, K. Watanabe, T. Taniguchi, Z. Mao, W. Windl, J. Goldberger, M. Bockrath, Ch. Lau, Raman Spectroscopy, Photocatalytic Degradation, and Stabilization of Atomically Thin Chromium Tri-iodide, Nano Lett. 18 (2018) 4214−4219.

[19] M. A. McGuire, Crystal and Magnetic Structures in Layered, Transition Metal Dihalides and Trihalides, Crystals 7 (2017) 121.

[20]  D. F. C. Morris, The instability of some dihalides of copper and silver, J. Phys. Chem. Solids 7 (1958) 214-217.

[21]  S. Kazim, S. Palleschi, G. D'Olimpio, D. Mastrippolito, A. Politano, R. Gunnella, A. Di Cicco, M. Reopt, G. Moccia, O. A. Cacioppo, R. Alfonsetti, J. Strychalska-Nowak, T. Klimczuk, R. J. Cava, L. Ottaviano, Mechanical exfoliation and layer number identification of single crystal monoclinic $CrCl_3$, Nanotechnology 31 (2020) 395706.

[22]  A. Mogus-Milankovic, J. Ravez, J. P. Chaminade, P. Hagenmuller, Ferroelastic properties of $TF_3$ compounds (T = Ti, V, Cr, Fe, Ga), Mat. Res. Bull., 20 (1985) 9-17.

[23]  K. Murata, H. Yoshino, H. O. Yadav, Y. Honda, N. Shirakawa, Pt resistor thermometry and pressure calibration in a clamped pressure cell with the medium Daphne 7373, Rev. Sci. Instrum. 68 (1997) 2490-2493.

[24]  K. Yokogawa, K. Murata, H. Yoshino, S. Aoyama, Solidification of High-Pressure Medium Daphne 7373, Jpn. J. Appl. Phys. 46 (2007) 3636-3639.

[25]  A. K. Kundu, Y. Liu, C. Petrovic, T. Valla, Valence band electronic structure of the van der Waals ferromagnetic insulators: $VI_3$ and $CrI_3$, Sci. Rep. 10 (2020) 15602.

[26]  F. Subhan, J. Hong, Magnetic anisotropy and Curie temperature of two-dimensional $VI_3$ monolayer, J. Condens. Matter Phys. 32 (2020) 245803.